\documentclass[journal,twoside,web]{ieeecolor}
\usepackage{tmi}
\usepackage{cite}
\usepackage{amsmath,amssymb,amsfonts}
\usepackage{algorithmic}
\usepackage{float}
\usepackage{graphicx}
\usepackage{textcomp}
\usepackage{multirow}
\usepackage{hyperref}

\begin{document}
\title{Two-Stream Graph Convolutional Network for Intra-oral Scanner Image Segmentation}
\author{\hspace*{0pt}Yue Zhao, 
Lingming Zhang,
Yang Liu,
Deyu Meng, \IEEEmembership{Member, IEEE},
Zhiming Cui,
Chenqiang Gao,
Xinbo Gao, \IEEEmembership{Senior Member, IEEE},
Chunfeng Lian, \IEEEmembership{Member, IEEE}, and
Dinggang Shen,\IEEEmembership{Fellow, IEEE}
\thanks{Yue Zhao, Lingming Zhang and Chenqiang Gao are with Chongqing University of Posts and Telecommunications, School of Communication and Information Engineering, and also with the Chongqing Key Laboratory of Signal and Information Processing, Chongqing, 400065, China.}
\thanks{Yang Liu is with Department of Orthodontics, Stomatological Hospital of Chongqing Medical University, and also with Chongqing Key Laboratory for Oral Diseases and Biomedical Sciences, Chongqing, 401147, China.}
\thanks{Deyu Meng and Chunfeng Lian are with Faculty of Information Technology, Macau University of Science and Technology, Macau, China and School of Mathematics and Statistics, Xi'an Jiaotong University, Xian, 710049, China.}
\thanks{Zhiming Cui and Dinggang Shen are with School of Biomedical Engineering, ShanghaiTech University, Shanghai, 201210, China. Dinggang Shen is also with Shanghai United Imaging Intelligence Co., Ltd., Shanghai, 200030, China. Zhiming Cui is also with the School of Computer Science, The University of Hong Kong, HK 999077, China.}
\thanks{Xinbo Gao is with School of Computer Science and Technology, Chongqing University of Posts and Telecommunications, Chongqing, 400065, China.}
\thanks{Corresponding authors: Dinggang Shen (e-mail: Dinggang.Shen@\\gmail.com) and Chenqiang Gao (e-mail: gaocq@cqupt.edu.cn).}
}
\maketitle

\begin{abstract}
 Precise segmentation of teeth from intra-oral scanner images is an essential task in computer-aided orthodontic surgical planning. 
 The state-of-the-art deep learning-based methods often simply concatenate the raw geometric attributes (i.e., coordinates and normal vectors) of mesh cells to train a single-stream network for automatic intra-oral scanner image segmentation. 
 However, since different raw attributes reveal completely different geometric information, the naive concatenation of different raw attributes at the (low-level) input stage may bring unnecessary confusion in describing and differentiating between mesh cells, thus hampering the learning of high-level geometric representations for the segmentation task. 
 To address this issue, we design a two-stream graph convolutional network (i.e., TSGCN), which can effectively handle inter-view confusion between different raw attributes to more effectively fuse their complementary information and learn discriminative multi-view geometric representations. 
 Specifically, our TSGCN adopts two input-specific graph-learning streams to extract complementary high-level geometric representations from coordinates and normal vectors, respectively. 
 Then, these single-view representations are further fused by a self-attention module to adaptively balance the contributions of different views in learning more discriminative multi-view representations for accurate and fully automatic tooth segmentation. 
 We have evaluated our TSGCN on a real-patient dataset of dental (mesh) models acquired by 3D intraoral scanners. 
 Experimental results show that our TSGCN signiﬁcantly outperforms state-of-the-art methods in 3D tooth (surface) segmentation.
 Github: \url{https://github.com/ZhangLingMing1/TSGCNet}.
\end{abstract}


\section{Introduction}
With the advancement in computer hardware and software technology, computer-aided-design (CAD) systems are being widely used by orthodontists, for signiﬁcantly improving treatment efﬁciency in modern dentistry. 
One essential task of an advanced CAD system is to perform fully automatic tooth segmentation on the intra-oral scanner images reconstructed by  the  intra-oral scanners (IOS). 
In this task, accurate labeling of each tooth and the derived information from labeled teeth are critical for various subsequent tasks towards precise personalized treatment, including diagnosis, patient-speciﬁc treatment planning, and treatment outcome evaluation.

However, segmenting teeth from intra-oral scanner image is challenging due to at least three reasons. 
1) Each tooth’s shape is unique and has large variation across individuals; 
2) Orthodontic patients (our target patients) often have atypical dental conditions, including missing, crowded, and misaligned teeth, which may result in complicated tooth boundaries; 
3) Teeth in deep intra-oral regions (e.g., the 2nd molar) may not be fully captured due to occlusion during scanning.

So far, there are two categories of conventional methods proposed to segment teeth from intra-oral scanner images. 
1) The first category of methods, i.e., projection-based methods~\cite{T9,T10}, usually first project the 3D intra-oral scanner image onto a 2D space to perform image-wise segmentation, and then reconstruct the segmentation result back to the original 3D space.
Although straightforward, the accuracy of these projection-based methods is limited due to the loss of spatial information in 3D-to-2D projection.
2) The second category of methods, i.e., geometry-based methods~\cite{T2,T3,T4,T5,T6,T7,T8}, typically use pre-selected geometric attribute (e.g., 3D coordinates, normal vectors, and curvatures) to separate mesh cells. 
However, these geometry-based methods are not fully automatic, as manual initialization relying on domain knowledge and experience is often required. 
Besides, the low-level pre-deﬁned attributes used in these geometry-based methods are very sensitive to dramatic variation of tooth appearances in patients.

Encouraged by the successful applications of convolutional neural networks (CNNs) in computer vision and medical image computing, some CNN-based methods have also been proposed to segment teeth from intra-oral scanner images.
Considering that the general CNNs are restricted to process images with regular shapes, these CNN-based methods typically organize hand-crafted feature vectors as 2D images~\cite{dental2} or voxelize unordered mesh vertices/cells as 3D grid volumes~\cite{Dental1}, which are then used as the input of a segmentation network. 
Such operations inevitably ignore the unordered nature of geometric data (e.g., different hand-crafted features of a cell have no spatial relationship), or may introduce additional computational costs and quantization errors during voxelization, thus hampering the segmentation accuracy on 3D meshes.
Along with the advancements of end-to-end deep learning for 3D shape analysis~\cite{Pointnet}, more recent works proposed to learn translation-invariant geometric features from the raw mesh data for vertex/cell-wise labeling on 3D dental surfaces~\cite{dental3, dental4, dental5}.
Although efficient and have achieved state-of-the-art segmentation performance, these end-to-end deep learning methods often simply concatenate different raw attributes as the input vector to train a single-stream segmentation network, potentially resulting in isolated false predictions on the intra-oral scanner image.
This is mainly because different raw attributes, e.g., the coordinates (the cell spatial position) and normal vectors (the cell morphological structure), have completely different geometric meanings, due to which their native combination at the input stage may introduce unnecessary inter-view confusion, thus hampering the seamless fusion of their complementary information to learn high-level multi-view representations.

In this paper, we propose a two-stream graph convolutional network (i.e., TSGCN) to learn discriminative geometric features from heterogenous multi-view inputs for end-to-end tooth segmentation from 3D dental meshes.
Our TSGCN has two critical components.
1) It starts with two parallel branches consisting of input-specific graph-learning modules, which learn high-level single-view representations from the coordinates and normal vectors, respectively.
2) These complementary single-view representations are then combined by a fusion branch integrating a self-attention mechanism, which minimizes the inter-view confusion and adaptively balances the contributions of different inputs to learn high-level multi-view geometric representations for the segmentation task.
Our TSGCN has been evaluated on a real-patient dataset of intra-oral scanner images acquired by IOS, leading to superior tooth segmentation performance compared with the state-of-the-art end-to-end methods.

This work is a comprehensive extension of a preliminary conference paper~\cite{TSGCNet}.
 Compared with the preliminary version, the major extensions are three-fold. 1) The fusion stage of our TSGCN applies a mesh-wise  normalization to eliminate the numerical gap between the single-view features extracted in the two parallel streams.
 2) The fusion stage of our TSGCN also integrates a self-attention mechanism to adaptively balance the contributions of different inputs, which further enhances the discriminative power of the learned multi-view feature representations. 
 3) The performance of our TSGCN as well as the efﬁcacy of its key components have been systematically justiﬁed by more comprehensive ablation studies.

\begin{figure*}[t]
\begin{center}
\includegraphics[scale=0.98]{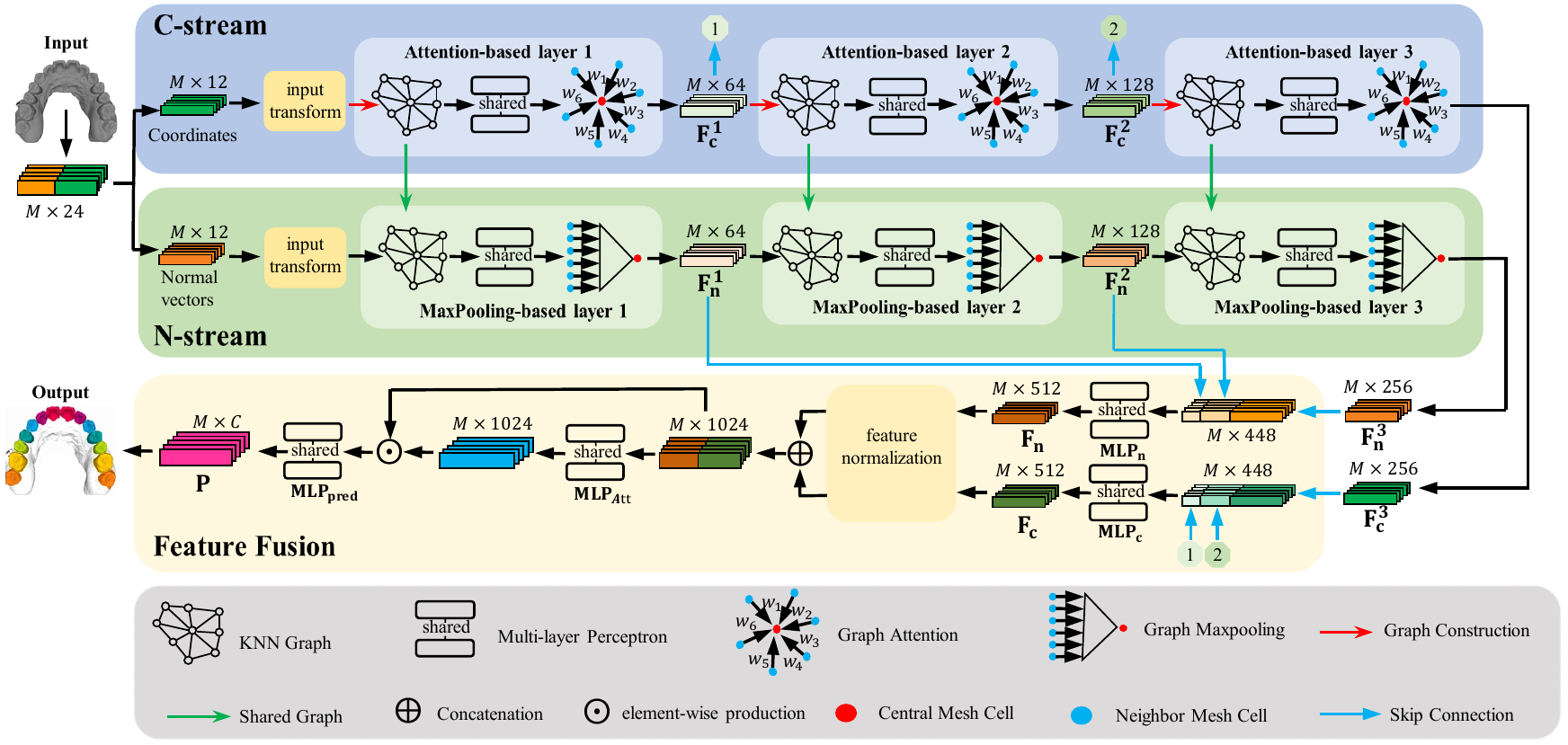}
\end{center}
\caption{Structure of our TSGCN. The network takes raw mesh data as inputs, and adopts two independent graph convolutional streams (i.e., C-stream and N-stream) to learn discriminative geometric representations from different features(i.e., 3D coordinates and normal vectors of meshes). Then, the high level features produced by two streams are fused for final mesh-wise tooth segmentation. Note that the circled numbers 1 and 2 denote the skip-connection, thus each same number in two different places will be connected together.}
\label{structure}
\end{figure*}

The rest of the paper is organized as follows.  The most related works, including 3D shape segmentation and intra-oral scanner image segmentation, are brieﬂy reviewed in Section~\ref{related}.
The studied data and our TSGCN are described in Section~\ref{method}.
Section~\ref{experiment} presents experimental results and comparisons of our TSGCN with other state-of-the-art methods. 
We discuss the effectiveness of each key module of our TSGCN  in Section~\ref{abalation}. 
Finally, the paper is concluded in Section~\ref{conclusion}. 

\section{Related Work}
\label{related}
In this section, we brieﬂy review existing methods in the literature that are closely related to our study, including those for general 3D shape segmentation as well as intra-oral scanner image segmentation. 
\subsection{3D Shape Segmentation}
Diverse deep learning methods have been proposed for  3D shape segmentation, which can be roughly grouped as 1) view-based, 2) voxel-based, 3) point-based, and 4) graph-based methods, as briefed below.

\subsubsection{View-based Methods}
View-based methods~\cite{PM1,PM2,PM3,PM4} typically project 3D geometric data (e.g., 3D point cloud) to 2D images based on predefined settings of view angles. 
Then, the projected images are processed by general CNNs to extract features. 
Such view-based methods have achieved promising performance in the shape classification task~\cite{PM6}. However, since the 2D projection inevitably results in spatial information loss, their performance in the dense segmentation task is limited.

\subsubsection{Voxel-based Methods}
Voxel-based methods~\cite{PV1,PV2,PV4,PV5,PV6,PV7,PV8} discretize/voxelize the 3D space into regular volumetric occupancy grids, after which regular 3D CNNs are applied to performing segmentation. 
Although straightforward, the volumetric representations incline to introduce quantization artifacts that hamper segmentation accuracy, as the time and space complexity heavily restrict the resolution of the volumetric representation.

\subsubsection{Point-based Methods}
Point-based methods aim to use deep learning architectures to directly process 3D geometric data. 
For example, PointNet~\cite{Pointnet} applied successive multi-layer perceptrons (MLPs) and a symmetric function (e.g., global max-pooling) to learn translation-invariant geometric features from irregular point clouds.
Although PointNet has achieved promising results in multiple tasks, it tends to ignore local spatial relationships on 3D shapes as its architecture learns features for each cell independently.
To address this limitation, PointNet++~\cite{Pointnet++} constructed a hierarchical architecture that recursively applies PointNet to exploit local spatial relationships on 3D shapes.
To learn more detailed local geometric information, other works further extended PointNet++ by integrating attention modules~\cite{AM1}, geometry sharing modules~\cite{geometry} and edge branches~\cite{EdgeBranch}.
Similarly, PointCNN~\cite{pointcnn} adopted an encoder-decoder architecture with $\chi$-transformations of unordered points to perform general convolutional operations.

\subsubsection{Graph-based methods} 
Recently, the graph CNNs have shown great success due to their flexibility in learning from non-Euclidean data. 
Many graph CNN-based methods have also been proposed for 3D shape recognition and segmentation~\cite{Att1,Att2}. 
They usually represented the 3D data as a graph according to the spatial relations between points/cells, and then used spectral-based~\cite{spectral1,spectral2,Att3} or spatial-based~\cite{spatial1} graph convolutions to aggregate local information for each node.

\subsection{Intra-oral Scanner Image Segmentation}
Conventional intra-oral scanner image segmentation methods based on pre-selected geometric properties can be roughly classified as curvature-based, contour-line-based, and harmonic-field-based methods.
Curvature-based methods~\cite{T1,T2,T3} usually leverage the negative curvature features to divide the surface into different parts.
For example, Yuan et al.~\cite{T1} classified different regions of intra-oral scanner images based on the minimum curvatures of the surface.
Zhao et al.~\cite{T2} proposed an interactive segmentation method based on curvature values of the triangle mesh.
Contour-line-based methods~\cite{T6,T7} allow the human interaction during segmentation to improve performance.
Specifically, users can initialize the boundary between each tooth and gum, and then the algorithm connects each pair of the neighboring points depending on the geodesic information.
Harmonic-field-based methods~\cite{T8} require users to annotate a limited number of surface points as prior and subsequently employ a harmonic field to segment the tooth.
Due to the need of specialized domain knowledge and human operations, the efficacy of such semi-automated methods heavily depends on the expertise of an operator.

Recently, several deep learning-based methods have been proposed for fully automated tooth segmentation~\cite{x-ray1,x-ray2}. 
For example, Xu et al.~\cite{dental2} proposed to reshape cell-wise hand-crafted geometric features as 2D image patches to train 2D CNNs for classifying the mesh cells.
Tian et al.~\cite{Dental1} first voxelized the intra-oral scanner image with a sparse octree partitioning, and then applied standard 3D CNNs for tooth segmentation.
However, converting the intra-oral scanner image into grid format in these methods tends to ignore the unordered nature of the geometric data~\cite{dental2} or may introduce additional quantization errors during the voxelization step~\cite{Dental1}.
Inspired by the success of the point-cloud segmentation networks, 
Zanjani et al.~\cite{dental3} proposed an end-to-end network that integrates PointCNN~\cite{pointcnn} with a discriminator to directly segment the raw dental surfaces acquired by IOS. 
Lian et al.~\cite{dental5} extended PointNet~\cite{Pointnet} by adding a multi-scale graph-constrained module to extract fine-grained local geometric features from dental mesh data.
Based on PointNet++~\cite{Pointnet++}, Cui et al.~\cite{TSegNet} proposed a two-stage algorithm to perform tooth segmentation, in which the first stage is to detect all the teeth and the second stage is to segment each tooth.
Instead of solely using the 3D coordinates, these deep-learning methods combined 3D coordinates and normal vectors as the network inputs and designed a single-stream network architecture for segmentation.
However, since coordinates and normal vectors have completely different geometric interpretations of a 3D shape, directly combinging  the mixed geometric inputs could confuse these single-stream networks during the learning of discriminative multi-view representations.
Different from those methods, our TSGCN adopts two graph-learning streams to independently learn feature representations from coordinates and normal vectors. In this way, the mutual confusion caused by mixed geometric inputs can be largely eliminated.

\section{Materials and Method}
\label{method}
\subsection{Data and Pre-processing}
The studied dataset consists of 80 intra-oral scanner images acquired by an IOS (Invisalign iTero) from different orthodontic patients. 
Each raw intra-oral scanner image contains approximately more than 100,000 mesh cells, which were downsampled to 16,000 mesh cells by preserving the original topology. 
As the input of our TSGCN, each cell of a downsampled mesh is described by a $24$-dimensional vector, including the 3D coordinates ($12$ elements) and normal vectors ($12$ elements) of the cell’s three vertices and its central point. 
That is, the network input is an $M\times24$ matrix (e.g., $M=16,000$). 

\subsection{Structure of TSGCN}
\label{Two Stream}
\subsubsection{Overview} 
As illustrated in Fig.~\ref{structure}, our TSGCN starts with two parallel streams, i.e., the C-stream and N-stream, which adopt input-specific graph-learning layers to extract high-level geometric representations from the coordinates and normal vectors, respectively. 
After that, these single-view features produced by these two complementary streams are further combined in the feature-fusion branch to learn more discriminative multi-view representations for teeth segmentation.  
Briefly, given the input of an $M\times 24$ matrix of cell-wise raw attributes,  our TSGCN outputs an $M\times C$ matrix, with each row denoting the probabilities of the respective cell belonging to $C$ different classes.

\subsubsection{C-Stream} 
Our C-stream is designed to capture the basic topology of an intra-oral scanner image from coordinates  of all cells. 
Given the input of an $M\times 12$ coordinate matrix $\mathbf{F}_\mathbf{c}^{0}$, the C-stream first adopts an input-transformer module (consisting of MLPs shared across cells) to learn an affine transformation matrix $\mathbf{T} \in \mathbb{R}^{12\times 12}$, which updates $\mathbf{F}_\mathbf{c}^{0}$ as:
\begin{equation}
\label{transform}
\hat{\mathbf{F}}_\mathbf{c}^{0} = \mathbf{F}_\mathbf{c}^{0} \mathbf{T}.
\end{equation}
In this way, the network inputs of different intra-oral scanner images can be aligned to a canonical space, which stabilizes the extraction of more representative geometric features in the subsequent layers ~\cite{Pointnet}.

Following the input-transformer module, a series of graph-attention layers are successively applied in the forward path of the C-stream to hierarchically extract multi-scale geometric features from the coordinate aspect. 
Specifically, given the feature matrix $\mathbf{F}^{l}_{c}\in \mathbb{R}^{M \times d}$ learned by $(l-1)$-th graph-attention layer, where the row vector $\mathbf{f}^{l}_{i}\in \mathbb{R}^d$ denotes the representation of the $i$-th cell $m_i$, the subsequent $l$-th graph-attention layer further extracts high-level geometric representations $\mathbf{F}^{l+1}_{c}\in \mathbb{R}^{M\times k}$ in four steps.

\textit{First}, in terms of $\mathbf{F}^{l}_{c}$,  we construct a dynamic KNN graph $G(V,E)$, where $V=\{m_1,m_2,...,m_M\}$ and $E\subseteq| V|\times|V|$ denote the set of $M$ nodes (mesh cells) and the corresponding set of edges (defined by the KNN connectivity), respectively.
Notably, each node  $m_i\in V$ only connects to its KNNs, which can be denoted as $\mathcal{N}(i)$.

\textit{Second}, we calibrate the local information for each center $m_i$.
That is, the representation $\mathbf{f}^{l}_{ij}$ of the $j$-th nearest neighbor $m_{ij}\in\mathcal{N}(i)$ is updated for $m_i$ via integrating its own representation, such as:
\begin{equation}
\label{update}
\hat{\mathbf{f}}^{l}_{ij}=\mathbf{MLP}^{l}\Bigl(\mathbf{f}^{l}_{i} \oplus \mathbf{f}^{l}_{ij}\Bigl),\,\, \forall\, m_{ij}\in \mathcal{N}(i),
\end{equation}
where $\oplus$ indicates the channel-wise concatenation, and $\hat{\mathbf{f}}^{l}_{ij}\in \mathbb{R}^{k}$ is the calibrated neighborhood representation.
In this way, the information provided by $m_{ij}$ (encoded in $\hat{\mathbf{f}}^{l}_{ij}$) can be more consistent with the node $m_i$, considering that $m_{ij}$ could be a nearest neighbor of more than one centers on $G(V,E)$, i.e., $\mathbf{f}^{l}_{ij}$ might be shared by multiple nodes.

\textit{Third}, we estimate the attention weights for the neighborhood $\mathcal{N}(i)$ of each node $m_i$.
Inspired by \cite{Att1,Att2}, such attention weights in this work are learned in a task-oriented fashion by using a lightweight network shared across cells/nodes, which can flexibly capture local geometric characteristics of an intra-oral scanner image for the segmentation task.
Specifically, the attention weight $\mathbf{\alpha}^{l}_{ij} \in \mathbb{R}^k$ of neighbor $m_{ij}$ in the $l$-th layer is defined as:
\begin{equation}
\label{weight}
\mathbf{\alpha}^{l}_{ij}=\sigma\Bigl(\Delta \mathbf{f}^{l}_{ij}\oplus \mathbf{f}^{l}_{ij}\Bigl), \,\,\forall\, m_{ij}\in \mathcal{N}(i),
\end{equation}
where the function $\sigma(\cdot)$ is implemented as a lightweight MLP, which adopts both $\Delta \mathbf{f}^{l}_{ij} = \mathbf{f}^{l}_i-\mathbf{f}^{l}_{ij}$ and $\mathbf{f}^{l}_{ij}$ as the inputs.
In the input feature space, the $\Delta \mathbf{f}^{l}_{ij}$ quantifies the dissimilarity between $m_{i,j}$ and $m_i$, which guides the current layer to assign more attention to closer neighbors of the center $m_i$; on the other hand, the $\mathbf{f}^{l}_{ij}$ provides detailed neighbor information of $m_{i,j}$. 

\textit{Finally}, we aggregate the neighborhood information to each center, which is formulated as:
\begin{equation}
\label{aggregate}
\mathbf{f}^{l+1}_{i} = \sum_{m_{ij}\in \mathcal{N}(i)} \mathbf{\alpha}^{l}_{ij} \odot \hat{\mathbf{f}}^{l}_{ij},
\end{equation}
where $\odot$ performs the element-wise production of two feature vectors, and the output $\mathbf{f}^{l+1}_{i}$ indicates the updated feature representation of $m_i$, i.e., the input feature of the $(l+1)$-th layer.
Here, $\mathbf{\alpha}^{l}_{ij}$ and $\hat{\mathbf{f}}^{l}_{ij}$ are defined by Eq. (\ref{weight}) and Eq. (\ref{update}), respectively.

\subsubsection{N-Stream} 
Although the C-stream can learn the basic structure of an intra-oral scanner image from the cells’ coordinates, it cannot sensitively distinguish adjacent cells belonging to different classes (e.g., teeth boundaries).
As a complementary branch to the C-stream, we further design an N-stream to learn fine-grained boundary representations from the aspect of normal vectors.

Our N-stream takes as inputs the normal vectors for all cells, which are aligned by an input-transformer module to a canonical space before the hierarchical extraction of higher-level feature representations.
To learn boundary representations in local regions and avoid the disturbance between distant cells with similar normal vectors (but belonging to different classes), the N-stream is restricted to share the same KNN graphs constructed in the C-stream.
In contrast to the case of using the same KNN graphs, the N-stream adopts graph max-pooling layers different from the graph-attention layers in the C-stream for feature extraction, mainly considering that the normal vectors reveal completely different geometric information compared with the coordinates.

Specifically, we assume $\mathbf{F}^{l}_{n}\in \mathbb{R}^{M \times d}$ is the input feature matrix of the $l$-th graph max-pooling layer in the N-stream.For simplicity, we still use the symbol $\mathbf{f}^{l}_{i}$ to denote the feature vector of a node $m_i$  (i.e., the $i$-th row of $\mathbf{F}^{l}_{n}$), and the corresponding feature of its neighbor $m_{ij} \in \mathcal{N}(i)$ is $\mathbf{f}^{l}_{ij}$.
The $l$-th graph max-pooling layer first calibrates the local information for each node $m_i$, by updating $\mathbf{f}^{l}_{i}$ as $\hat{\mathbf{f}}^{l}_{i}$ according to Eq.~(\ref{update}).
Thereafter, the channel-wise max-pooling is further applied on all neighbors’ calibrated features to produce the boundary representation for the respective center $m_i$, which can be formulated as:
\begin{equation}
\label{maxpooling}
\mathbf{f}^{l+1}_{i} = maxpooling\Bigl\{\hat{\mathbf{f}}^{l}_{ij},\,\, \forall \,m_{ij}\in \mathcal{N}(i)\Bigl\}.
\end{equation}
It is worth mentioning that we use max-pooling (rather than graph attention) in the N-steam since the max operator can more sensitively capture the most distinctive features presented at the tooth boundaries.

\subsubsection{Feature Fusion}
\label{Feature FUsion}
As shown in Fig.~\ref{structure}, after extracting single-view representations in the C-stream and N-stream, respectively, our TSGCN further fuses them to learn more discriminative multi-view representation for teeth segmentation.
To this end, the multi-scale cell-wise features from different layers in each stream (i.e., $\mathbf{F}^{l}_{c}$ or $\mathbf{F}^{l}_{n}$, where $l$ denotes the $l$-th layer) are concatenated, on which an MLP (i.e., MLP$\mathbf{_c}$ or MLP$\mathbf{_n}$) is applied to learn high-level single-view representations (i.e., $\mathbf{F_{c}}$ or $\mathbf{F_{n}}$) encoding the local-to-global information for the corresponding view (i.e., the C-stream or the N-stream). This operation can be formulated as:
\begin{equation}
\label{skip1}
\mathbf{F_c}=\mathbf{MLP}_\mathbf{c}\Bigl(\mathbf{F}^{1}_\mathbf{c} \oplus \mathbf{F}^{2}_\mathbf{c} \oplus \mathbf{F}^{3}_\mathbf{c}\Bigl),
\end{equation}

\begin{equation}
\label{skip2}
\mathbf{F_n}=\mathbf{MLP}_\mathbf{n}\Bigl(\mathbf{F}^{1}_\mathbf{n} \oplus \mathbf{F}^{2}_\mathbf{n} \oplus \mathbf{F}^{3}_\mathbf{n}\Bigl).
\end{equation}
After that, the single-view representations ($\mathbf{F_{c}}$ and $\mathbf{F_{n}}$) are further harmonized by using a mesh-wise normalization operation. Specifically, for the same row in $\mathbf{F_{c}}$ and $\mathbf{F_{n}}$, e.g., $\mathbf{f}_{\mathbf{c}}^{i} \in \mathbf{F}_\mathbf{c}$ and $\mathbf{f}_{\mathbf{n}}^{i} \in \mathbf{F}_\mathbf{n}$, the respective normalization factors are computed as:
\begin{equation}
\label{factor_coor}
\delta_{\mathbf{c}} = \frac{|\mathbf{f}_{\mathbf{n}}^{i}|}{|\mathbf{f}_{\mathbf{c}}^{i}| + |\mathbf{f}_{\mathbf{n}}^{i}|},
\end{equation}

\begin{equation}
\label{factor_nor}
\delta_{\mathbf{n}} = \frac{|\mathbf{f}_{\mathbf{c}}^{i}|}{|\mathbf{f}_{\mathbf{c}}^{i}| + |\mathbf{f}_{\mathbf{n}}^{i}|}.
\end{equation}
Then, the feature vectors $\mathbf{f}_{\mathbf{c}}^{i}$ and $\mathbf{f}_{\mathbf{n}}^{i}$ can be updated as:
\begin{equation}
\label{normalize_coor}
\hat{\mathbf{f}_{\mathbf{c}}^{i}} = \delta_{\mathbf{c}} \mathbf{f}_{\mathbf{c}}^{i},
\end{equation}

\begin{equation}
\label{normalize_nor}
\hat{\mathbf{f}_{\mathbf{n}}^{i}} = \delta_{\mathbf{n}} \mathbf{f}_{\mathbf{n}}^{i}.
\end{equation}
Such a normalization operation defined in Eqs.~(\ref{factor_coor})-(\ref{normalize_nor}) helps eliminate the numerical gap between the single-view features extracted in the two parallel streams.

However, the normalized single-view features $\hat{\mathbf{f}_{\mathbf{c}}^{i}}$ and $\hat{\mathbf{f}_{\mathbf{n}}^{i}}$ may still have mismatches. 
This is mainly because the N-stream of our TSGCN uses the graph max-pooling to aggregate local information, due to which $\hat{\mathbf{f}_{\mathbf{n}}^{i}}$ might be numerically stronger than
$\hat{\mathbf{f}_{\mathbf{c}}^{i}}$, thus probably resulting in unnecessary biases in the subsequent cross-view fusion. 
To address this issue, we adopt a self-attention mechanism to enable the network to adaptively balance the contributions of $\hat{\mathbf{f}_{\mathbf{c}}^{i}}$ and $\hat{\mathbf{f}_{\mathbf{n}}^{i}}$.
Specifically, for each cell $m_i$, the self-attention weight $\mathbf{\beta}_{i}$ is defined as:
 \begin{equation}
\label{self-attention}
\mathbf{\beta}_{i} = \mathbf{MLP}_\mathbf{Att} \Bigl( \hat{\mathbf{f}_{\mathbf{c}}^{i}} \oplus \hat{\mathbf{f}_{\mathbf{n}}^{i}} \Bigl),
\end{equation}
where MLP$_\mathbf{Att}$ is a lightweight MLP, and its output $\mathbf{\beta}_{i}$ has the same size as $(\hat{\mathbf{f}_{\mathbf{c}}^{i}} \oplus \hat{\mathbf{f}_{\mathbf{n}}^{i})}$.
The multi-view feature geometric representation of $m_i$ can be quantified as:
 \begin{equation}
\label{select feature}
\hat{\mathbf{f}^{i}} = \mathbf{\beta}_{i} \odot \Bigl( \hat{\mathbf{f}_{\mathbf{c}}^{i}} \oplus \hat{\mathbf{f}_{\mathbf{n}}^{i}} \Bigl).
\end{equation}

Finally, a MLP (i.e., MLP$\mathbf{_{pred}}$) is applied on the multi-view feature matrix $\hat{\mathbf{F}} = ({\hat{\mathbf{f}^{1}}, \hat{\mathbf{f}^{2}},...,\hat{\mathbf{f}^{M}}})$ to output an $M\times C$ matrix $\mathbf{P}$, where each row denotes the probabilities of a specific cell belonging to $C$ different classes. 

\subsection{Implementation Details}
\subsubsection{Network Details}
As shown in Fig.~\ref{structure}, the TSGCN contains a C-stream, an N-stream, and a feature-fusion branch. 
For both the C-stream and the N-stream, the MLPs in the first layer to the third layer contain one 1D Conv with 64 channels, 128 channels, and 256 channels, respectively.
The number $K$ of each KNN graph is set as 32. 
The graph attention function $\sigma(\cdot)$ is implemented as a MLP, which is followed by the channel-wise softmax to normalize the output weights.
In the feature-fusion part, both MLP$\mathbf{_c}$ and MLP$\mathbf{_n}$ contain a 1D Conv with 512 channels, MLP$_\mathbf{Att}$ contains a 1D Conv with 1024 channles, and MLP$\mathbf{_{pred}}$ contains four successive 1D Convs, each with 512, 256, 128, and $C$ channels, respectively. 
All 1D Convs are followed by the batch normalization and LeakyReLU, except the last one in MLP$\mathbf{_{pred}}$, which is followed by a tensor-reshape operation to output the $M \times C$ probability matrix.

\subsubsection{Training Details}
Our TSGCN was trained by minimizing the cross-entropy segmentation loss on two NVIDIA GTX 1080 GPUs for 200 epochs. We use the Adam optimizer with the mini-batch size setting as 4. The initial learning rate was 1e-3, which was reduced by 0.5 decay for every 20 epochs. 

\begin{figure}[H]
\begin{center}
\includegraphics[scale=0.6]{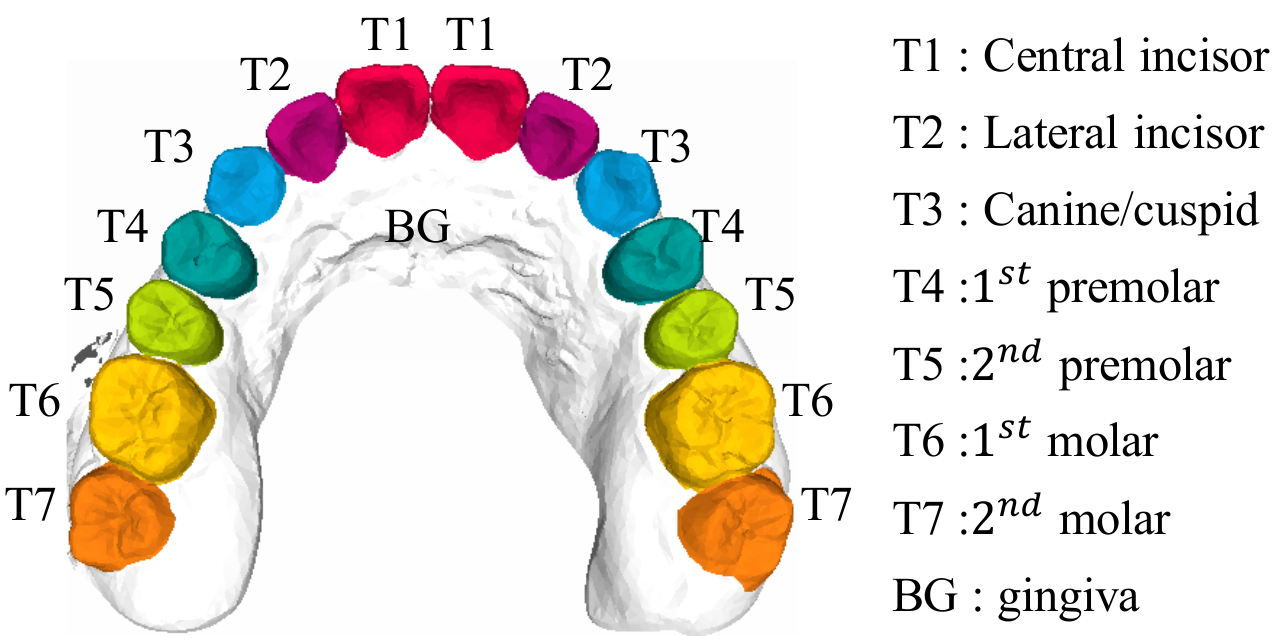}
\end{center}
   \caption{Illustration of a manually labeled intra-oral scanner image, 8 classes of teeth, i.e., the symmetric central incisor, lateral incisor, canine, $1^{st}$ premolar, $2^{nd}$ premolar, $1^{st}$ molar, $2^{nd}$ molar, and the gingiva.}
\label{label}
\end{figure}

\begin{table*}[]
\tiny
\centering
\caption{The segmentation results for five competing methods and our method on OA and mIoU.}
\label{result}
\resizebox{145mm}{16mm}{%
\begin{tabular}{c|c|c|cccccccc}
\hline
\multirow{2}{*}{Method} & \multicolumn{2}{c|}{All teeth} & \multicolumn{8}{c}{Each class (IoU)}  \\ \cline{2-11}                                                                                                                          
& OA            & mIoU           & \multicolumn{1}{c|}{T1} & \multicolumn{1}{c|}{T2} & \multicolumn{1}{c|}{T3} & \multicolumn{1}{c|}{T4} & \multicolumn{1}{c|}{T5} & \multicolumn{1}{c|}{T6} & \multicolumn{1}{c|}{T7} & BG  \\ \hline
PointNet\cite{Pointnet}                      & 84.95             & 66.86              & \multicolumn{1}{l|}{55.31}  & \multicolumn{1}{l|}{65.31}  & \multicolumn{1}{l|}{69.35}  & \multicolumn{1}{l|}{75.47}  & \multicolumn{1}{l|}{72.21}  & \multicolumn{1}{l|}{66.18}  & \multicolumn{1}{l|}{74.71} & 84.86  \\
PointCNN\cite{pointcnn}                      & 88.61             & 72.86              & \multicolumn{1}{l|}{61.72}  & \multicolumn{1}{l|}{66.45}  & \multicolumn{1}{l|}{68.10}  & \multicolumn{1}{l|}{78.98}  & \multicolumn{1}{l|}{78.57}  & \multicolumn{1}{l|}{70.51}  & \multicolumn{1}{l|}{72.15} & 86.39  \\
PointNet++\cite{Pointnet++}                      & 90.25             & 78.14              & \multicolumn{1}{l|}{67.82}  & \multicolumn{1}{l|}{74.61}  & \multicolumn{1}{l|}{78.10}  & \multicolumn{1}{l|}{82.73}  & \multicolumn{1}{l|}{80.70}  & \multicolumn{1}{l|}{74.67}  & \multicolumn{1}{l|}{78.94} & 87.52  \\
DGCNN\cite{spatial1}                      & 91.93             & 84.30              & \multicolumn{1}{l|}{82.18}  & \multicolumn{1}{l|}{79.95}  & \multicolumn{1}{l|}{82.09}  & \multicolumn{1}{l|}{87.88}  & \multicolumn{1}{l|}{86.24}  & \multicolumn{1}{l|}{80.14}  & \multicolumn{1}{l|}{84.26} & 91.65  \\
MeshSegNet\cite{dental4}                      & 93.11             & 84.47              & \multicolumn{1}{l|}{81.31}  & \multicolumn{1}{l|}{83.65}  & \multicolumn{1}{l|}{82.15}  & \multicolumn{1}{l|}{82.87}  & \multicolumn{1}{l|}{84.81}  & \multicolumn{1}{l|}{81.93}  & \multicolumn{1}{l|}{87.10} & 91.94  \\
Ours                      & \textbf{96.96}             & \textbf{91.69}              & \multicolumn{1}{l|}{\textbf{83.47}}  & \multicolumn{1}{l|}{\textbf{91.29}}  & \multicolumn{1}{l|}{\textbf{93.53}}  & \multicolumn{1}{l|}{\textbf{94.84}}  & \multicolumn{1}{l|}{\textbf{93.14}}  & \multicolumn{1}{l|}{\textbf{90.26}}  & \multicolumn{1}{l|}{\textbf{91.32}} & \textbf{95.67}  \\ \hline
\end{tabular}%
}
\end{table*}

\section{Experimental Results}
\label{experiment}
\subsection{Experimental Setup}
The task in this paper is to automatically segment each intra-oral scanner image as $C$ = 8 different semantic parts, including central incisor (T1), lateral incisor (T2), canine/cuspid (T3), 1$^{\text{st}}$ premolar (T4), 2$^{\text{nd}}$ premolar (T5), 1$^{\text{st}}$ molar (T6), 2$^{\text{nd}}$ molar (T7), and background/gingiva (BG). The ground-truth annotations of all intra-oral scanner images were defined according to the clinical requirement and professional dentists’ advice, with a typical example shown in Fig.~\ref{label}. The dataset was randomly split as a training set with 64 subjects, and a testing set with 16 subjects. Besides, we also augmented the training set by the combination of 1) random translation, and 2) random rotation of each intra-oral scanner image. 
Specifically, each training intra-oral scanner image was translated with a displacement randomly sampled between $[-10,10]$ and rotated along the $y$-axis with an angle randomly sampled between $[-\frac{\pi}{6}, \frac{\pi}{6}]$. 
In this way, we generated 64 new samples from each original intra-oral scanner image to enrich the diversity of the training set.

Our TSGCN is compared with five state-of-the-art methods for 3D shape segmentation (i.e., \textbf{PointNet}~\cite{Pointnet}, \textbf{PointNet}++~\cite{Pointnet++}, \textbf{PointCNN}~\cite{pointcnn}, \textbf{DGCNN}~\cite{spatial1}) and intra-oral scanner image segmentation (i.e., \textbf{MeshSegNet}~\cite{dental4}). 
The overall segmentation performance is quantitatively evaluated by two metrics, i.e., 1) Overall Accuracy (OA), which is calculated as: Nc (Number of correctly segmented cells) / N (Number of all cells), and 2) mean Intersection-over-Union (mIoU). Besides, we also calculate the detailed IoU of each class. 

\begin{figure*}[t]
\begin{center}
\includegraphics[scale=0.96]{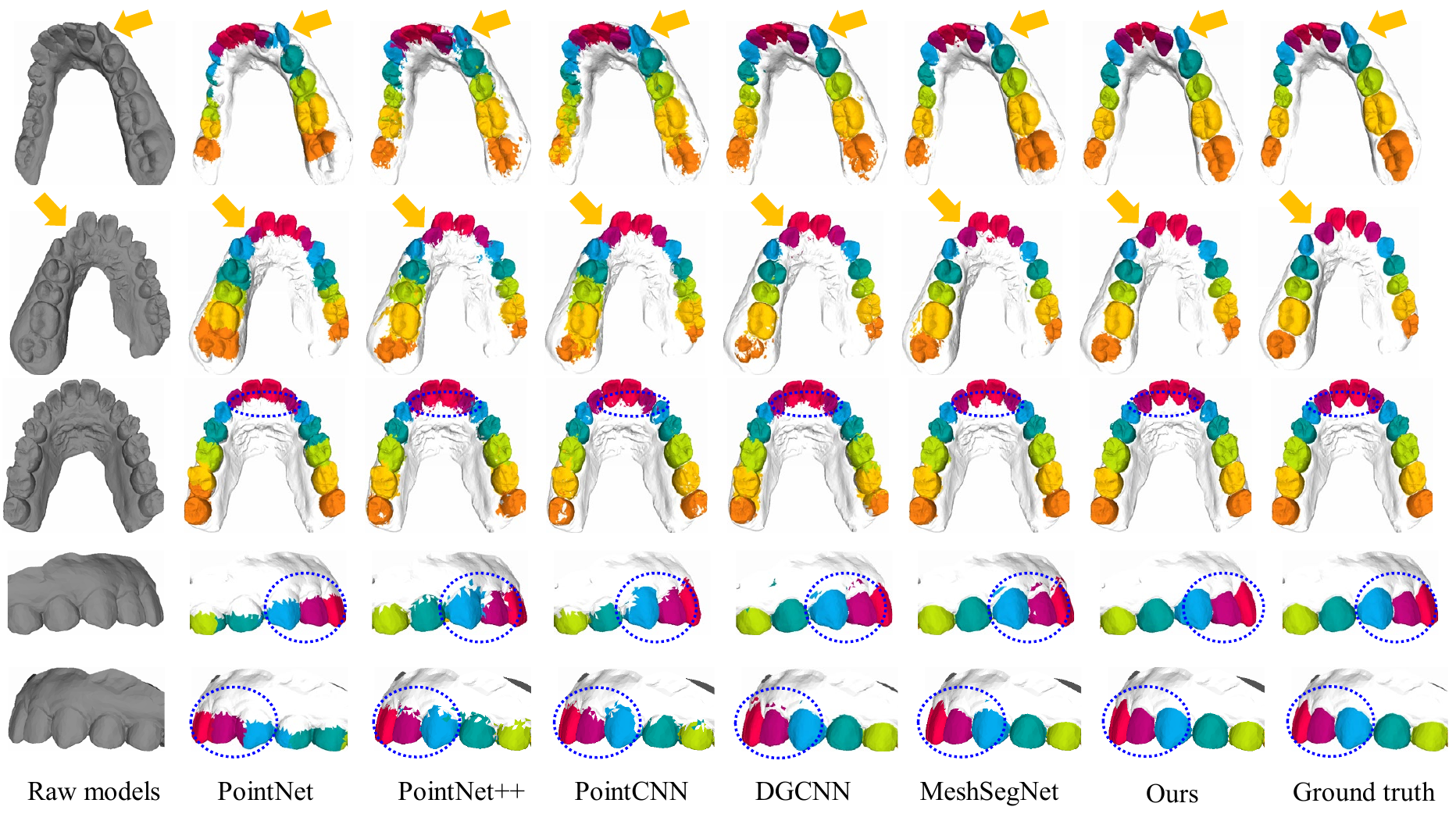}
\end{center}
   \caption{Visualization of representative segmentation results produced by five competing methods and our method, along with the respective ground-truth annotations.}
\label{segmental result}
\end{figure*}

\subsection{Comparison with Competing Methods}
The quantitative segmentation results obtained by all competing methods in terms of both OA and mIoU metrics are summarized in Table~\ref{result}.
From Table~\ref{result}, we can have at least four observations.
1) Our TSGCN consistently obtained superior overall accuracy than all the competing methods in terms of OA and mIoU, demonstrating the state-of-the-art performance by our TSGCN in automatic teeth segmentation.
2) Our TSGCN outperformed the MeshSegNet method~\cite{dental4} tailored for intra-oral scanner image processing, by improving OA and mIoU values for $3.85\%$ and $7.22\%$, respectively.
This suggests effectiveness of our network design in eliminating the inter-view confusion between coordinates and normal vectors, as MeshSegNet simply combines these raw attributes at the low-level input stage, while our TSGCN fuses them in the discriminative high-level feature space.
3) Our TSGCN signiﬁcantly outperformed the single-stream graph network DGCNN~\cite{spatial1}, demonstrating effectiveness of the proposed two-stream structure in learning discriminative geometric feature representations.
4) Our TSGCN consistently obtained better IoU values than other competing methods in segmenting each tooth, suggesting the generalization ability of our method in handling the varying teeth appearances.

Fig.~\ref{segmental result} presents the segmentation results of five representative intra-oral scanner images, from which we can have three observations. 
1) Consistent with quantitative results shown in Table~\ref{result}, our TSGCN can also qualitatively outperform all the competing methods.
Speciﬁcally, PointNet~\cite{Pointnet}, PointNet++~\cite{Pointnet++}, and PointCNN~\cite{pointcnn} failed to learn comprehensively from the complicated shape of the intra-oral scanner images,thus resulting in under-segmentation or over-segmentation for the misaligned teeth (as indicated by yellow arrows in the ﬁrst two rows).
Although graph-based competing methods (i.e., DGCNN~\cite{spatial1} and MeshSegNet~\cite{dental4}) achieved better performance based on the extraction of detailed local spatial information, they still failed to capture the complete tooth structure, mainly due to utilization of a single-stream architecture that cannot fully capture the complementary information from different raw attributes.
In contrast, by using the input-specific C-stream and N-stream, our TSGCN achieved more accurate results than all the competing methods in these misaligned areas.
2) From the third to the ﬁfth rows of Fig.~\ref{segmental result}, we can see that our TSGCN can also better segment boundaries between adjacent teeth, especially for the two adjacent incisors (as indicated by blue dotted circles), which demonstrates the effectiveness of our N-stream in learning distinctive structural details to distinguish tooth boundaries.
3) When comparing our method with DGCNN~\cite{spatial1} and MeshSegNet~\cite{dental4} in the fourth row, we can see that these two competing methods produced many isolated false predictions on the gingiva, even those mislabeled mesh cells are relatively far away from the real tooth area.
This further suggests that the direct concatenation of coordinates and normal vectors as a single feature vector (e.g., in MeshSegNet) may hamper the learning of discriminative geometric features in some cases, while the two-stream structure (i.e., in our TSGCN) is a more appropriate design.

\section{Discussion}
\label{abalation}
In this section, we conduct detailed ablation studies to evaluate the efficacy of the critical components of our TSGCN. 
We also discuss the limitations of our TSGCN and the potential solutions in the future.

\subsection{Effectiveness of the Two-Stream Structure}
\label{5.1}
Rather than the simple concatenation of raw geometric attributes at the input stage, our TSGCN adopts two parallel branches (C-stream and N-stream) to learn high-level single-view feature representation from the coordinates and normal vectors and then fuses their complementary information based on a self-attention mechanism.
In the subsequent series of experiments, we evaluate the effectiveness of our two-stream structure. 
Specifically, we remove the N-stream (i.e., only adopting the C-stream with the coordinates as input) or the C-stream (i.e., only adopting the N-stream with the normal vectors as input) to generate two different variants of our TSGCN, which are denoted as \textbf{TSGCN-C} and \textbf{TSGCN-N}, respectively. In addition, we also build another single-stream variant of TSGCN (denoted as \textbf{TSGCN-S}) that directly learns from the combination of coordinates and normal vectors. 
Note that TSGCN-S has a similar structure to TSGCN-C but with different input. 

\begin{table}[H]
\caption{The segmentation results for the original TSGCN and three variants. TSGCN-C and TSGCN-N stand for the sole use of the C-stream and N-stream, respectively. TSGCN-S denotes the single-stream version of TSGCN, which directly concatenates the coordinates and normal vectors as input.}
\begin{center}
\begin{tabular}{lll}
\hline
Structure           & OA    & mIoU  \\ \hline
TSGCN-C           & 83.23 & 63.79 \\
TSGCN-N           & 55.42 & 20.77 \\
TSGCN-S           & 87.25 & 73.44 \\
TSGCN             & \textbf{96.69} & \textbf{91.69} \\ \hline
\end{tabular}
\end{center}
\label{ablation1}
\end{table}

We compare these three variants with the final TSGCN, with the quantitative results listed in Table \ref{ablation1}. 
It can be seen that both TSGCN-N and TSGCN-C lead to worse results than both TSGCN-S and TSGCN. This justifies that the complementary geometric information provided by coordinates and normal vectors is significant for precise segmentation.
On the other hand, when compared with TSGCN-S, the original TSGCN further improves the segmentation accuracy. This suggests the effectiveness of our two-stream structure in extracting the discriminative geometric information from the two complementary but heterogeneous views. 

\subsection{Effectiveness of Feature-Aggregation Strategy}
As described in Section \ref{Two Stream}, we use two different feature aggregation strategies in the C-stream and the N-stream of our TSGCN. Specifically, the graph attention aggregation is used in the C-stream, while the graph max-pooling aggregation is used in the N-stream. 
To evaluate the effectiveness of our design, we implement three variants of TSGCN by changing the feature aggregation strategy in each stream, i.e., \textbf{1)} both streams use max-pooling, \textbf{2)} both streams use attention, and \textbf{3)} C-stream uses max-pooling while N-stream uses attention. For simplicity, we denote those three variants and the original TSGCN as \textbf{M+M}, \textbf{A+A}, \textbf{M+A}, and \textbf{A+M}, respectively. We compare segmentation results of these variants in Table \ref{ablation2}. From Table \ref{ablation2}, we can see that using attention mechanisms in the C-stream can achieve better performance (please refer to A+M  \emph{vs}. M+M) when compared with the case of using of max-pooling. This suggests that graph attention aggregation can capture fine-grained local geometric features of the tooth shape from coordinates. Besides, using max-pooling in the N-stream can further refine the segmentation results (please refer to A+M \emph{vs}. A+A). This can be rationally explained as: max-pooling can extract more distinctive morphological features, which in return helps the network capture difference between neighboring cells, especially at the tooth boundaries.

\begin{figure}[H]
\begin{center}
\includegraphics[scale=0.9]{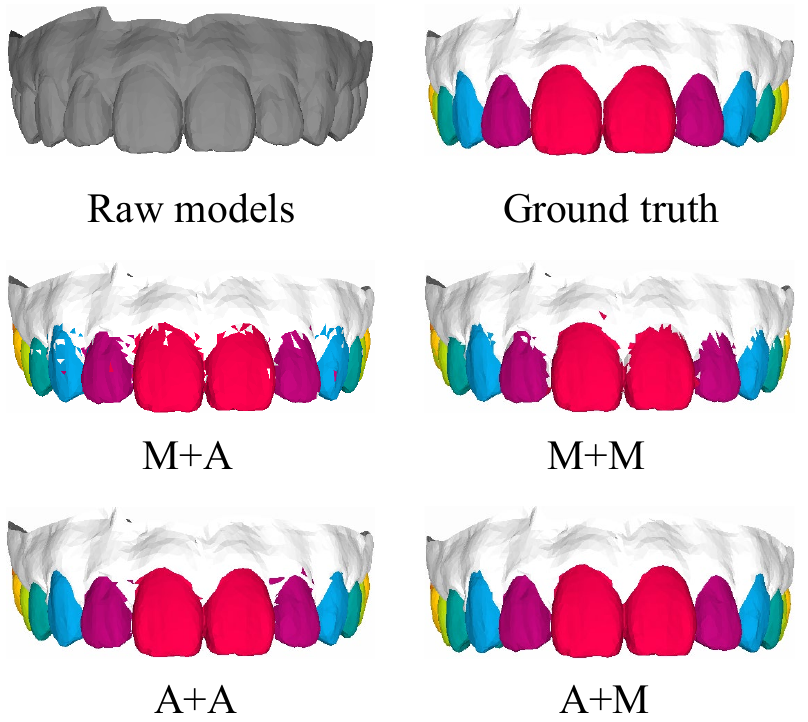}
\end{center}
   \caption{Segmentation example for TSGCN by using different feature aggregation strategies.}
\label{aggregation_example}
\end{figure}

We also show segmentation results of a typical example obtained by these variants in Fig.~\ref{aggregation_example}.
Consistent with the quantitative evaluations in Table \ref{ablation2}, we can see that both M+A and M+M have more outliers than A+A and A+M, which further confirm that graph attention aggregation is more suitable for the C-stream. Besides, when comparing A+A with A+M, we also observe that A+M generates more precise segmentation on boundaries, which further ascertains that graph max-pooling aggregation can facilitate the network to better distinguish the cells with similar coordinate information but belonging to different segmentation classes.

\begin{table}[H]
\caption{The segmentation results by using different feature aggregation strategies. M+M (or A+A) stands for using max-pooling (or attention) in both two streams. M+A stands for using max-pooling and attention in the C-stream and N-stream, respectively. A+M denotes the original TSGCN.}
\begin{center}
\begin{tabular}{lll}
\hline
Structure           & OA    & mIoU  \\ \hline
M+M           & 95.31 & 89.06 \\
A+A           & 96.52 & 91.19 \\
M+A           & 94.87 & 88.48 \\
A+M             & \textbf{96.69} & \textbf{91.69}\\ 
\hline
\end{tabular}
\end{center}
\label{ablation2}
\end{table}

\subsection{Effectiveness of Feature-Fusion Strategy}
Before the feature fusion part, the multi-scale high-level features produced by C-stream and N-stream (i.e., $\mathbf{F_{c}}$ and $\mathbf{F_{n}}$) are fused to learn complementary information. To evaluate the effectiveness of this high-level feature fusion strategy, we further compare TSGCN with another variant implemented by applying a low-level feature fusion strategy. Specifically, during the two-stream feature extraction stage, the output of the $l$-th layer in both streams are concatenated (i.e., $\mathbf{F^{l}_{c}}$ and $\mathbf{F^{l}_{n}}$ are concatenated) as the input of the $(l+1)$-th layer. This means that the C-stream and N-stream have the same input in the $(l+1)$-th layer. We denote our original feature fusion strategy and its variant as \textbf{H-fusion} and \textbf{L-fusion}, respectively. 

\begin{table}[H]
\caption{The segmentation results for two different feature fusion strategies. The L-fusion denotes low-level feature fusion strategy, and the H-fusion stands for our adopted feature fusion strategy.}
\begin{center}
\begin{tabular}{lll}
\hline
Strategy           & OA    & mIoU  \\ \hline
L-fusion           & 94.27 & 87.38 \\
H-fusion             & \textbf{96.69} & \textbf{91.69} \\ \hline
\end{tabular}
\end{center}
\label{ablation3}
\end{table}

Following this, we compared the segmentation results of H-fusion and L-fusion, as listed in Table \ref{ablation3}. From this table, it can be seen that the OA and mIoU of H-fusion are 2.42\% and 4.31\% higher than those of L-fusion, respectively. There is a possibility that the premature feature fusion also confuses the learning of discriminative features. Additionally, considering that some mesh cells have similar vector information but different coordinate information, the KNN graph built on the concatenated features may result in a random distribution of neighbors in real space, which tends to hamper the network to learn local-to-global information.

\subsection{Effectiveness of the Feature-Fusion Branch}
In the feature-fusion branch, we apply a mesh-wise feature normalization to eliminate the numerical gap between the single-view features from the C-stream and N-stream. After that, a self-attention mechanism is further applied to adaptively balance the contributions of different views in learning the multi-view feature representation. 
To evaluate the effectiveness of the above two designs, we alternatively used only the mesh-wise feature normalization or the self-attention mechanism in the feature-fusion branch, generating two variants of our TSGCN denoted as \textbf{TSGCN-Normalization} and \textbf{TSGCN-Attention}, respectively. 
Besides, the original TSGCN was also compared with another variant (denoted as \textbf{TSGCN-Concatenation}) that directly applies an MLP on the concatenation of the outputs of the C-stream and N-stream to perform feature fusion. 
The quantitative segmentation results obtained by these variants and the original network are compared in Table~\ref{ablation4}. We can see that both TSCGN-Normalization and TSGCN-Attention achieved better performance than TSGCN-Concatenation. This suggests that the mesh-wise feature normalization and self-attention mechanism successfully helped the network extract finer multi-view geometric features for more precise tooth segmentation. Moreover, by combining these two designs (i.e., the original TSGCN), the segmentation accuracy is further improved, which implies that both the mesh-wise feature normalization and the self-attention operations are important for the feature-fusion branch.

\begin{figure}[H]
\begin{center}
\includegraphics[scale=0.75]{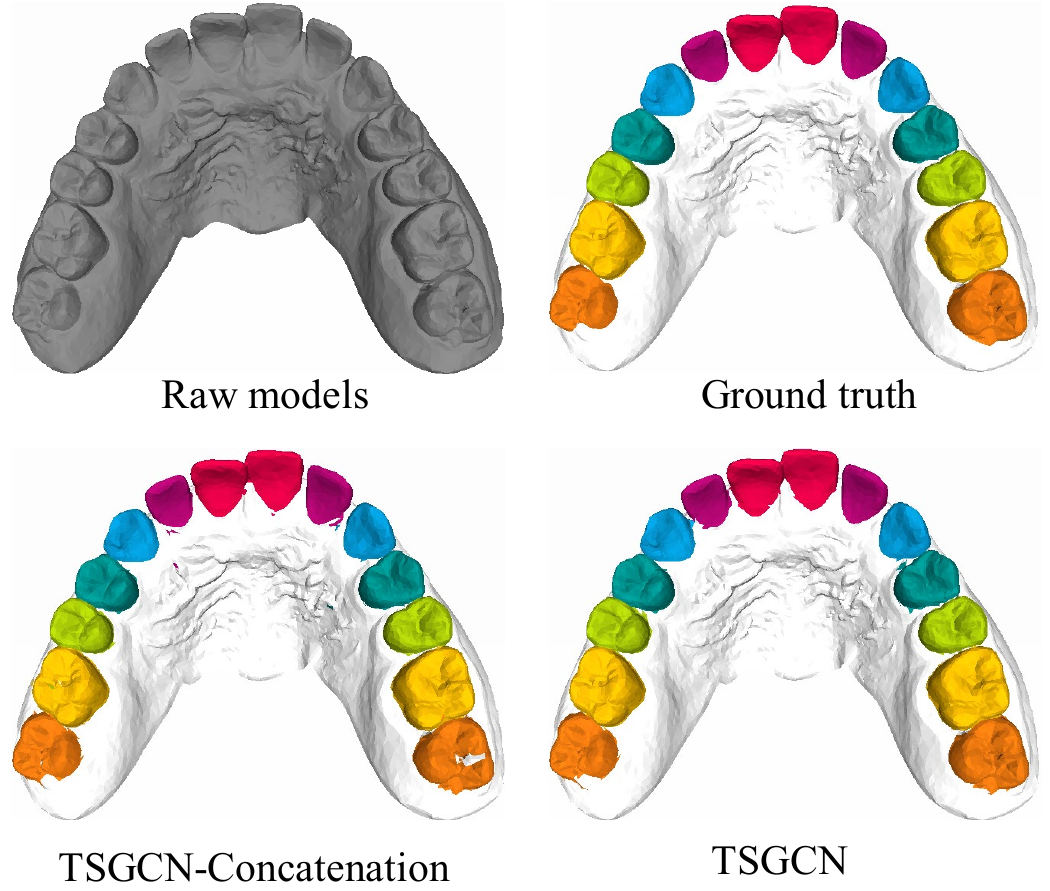}
\end{center}
\caption{A segmentation example for TSGCN by using different feature fusion structures.}
\label{feature fusion}
\end{figure}

We also visually compared the segmentation results obtained by TSGCN-Concatenation and TSGCN in Fig.~\ref{feature fusion}. From Fig.~\ref{feature fusion}, it is observed that TSGCN-Concatenation results in unexpected false predictions in the central area of $2^{nd}$ molar. 
One reason could be that the graph max-pooling operation in the N-stream may provide stronger geometric features related to normal vectors for feature fusion and this causes the network to classify the mesh cells with similar normal vectors as the same class directly, ignoring the position information provide by coordinates. 
In contrast, our TSGCN achieves more precise segmentation on the $2^{nd}$ molar, which further confirms the effectiveness of our proposed feature fusion structure in adaptively selecting multi-view features for more accurate segmentation.  

\begin{table}[H]
\caption{The segmentation results of different structure in feature fusion part. TSGCN-Normalization (or TSGCN-Attention) stands for only using mesh-wise feature normalization (or self-attention mechanism) in feature fusion. TSGCN-Concatenation stands for  directly applying MLP on the concatenation of two different view based features.}
\begin{center}
\begin{tabular}{lll}
\hline
Structure           & OA    & mIoU  \\ \hline
TSGCN-Concatenation           & 95.44 & 89.99 \\
TSGCN-Normalization           & 95.72 & 90.50 \\
TSGCN-Attention           & 96.16 & 90.95 \\
TSGCN             & \textbf{96.69} & \textbf{91.69}\\ 
\hline
\end{tabular}
\end{center}
\label{ablation4}
\end{table}

\subsection{Sensitivity to Different Numbers of Nearest Neighbors}
\begin{figure}[H]
\begin{center}
\includegraphics[scale=0.5]{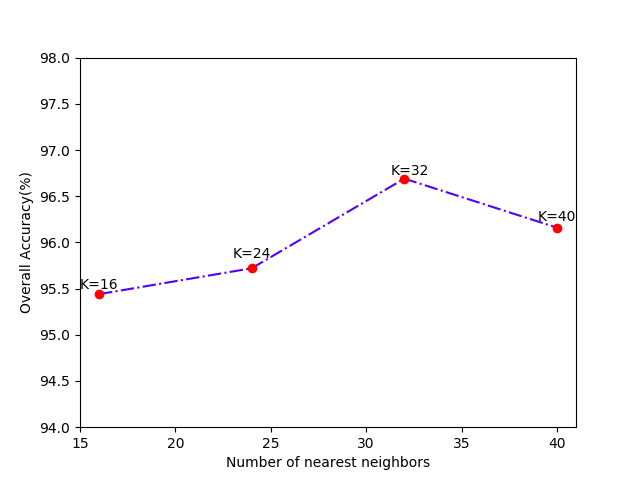}
\end{center}
\caption{The segmentation results of TSCGN with respect to different numbers $K$ of nearest neighbors in KNN graph.}
\label{ablation5}
\end{figure}

We also evaluated the sensitivity of our TSGCN with respect to different numbers of nearest neighbors in the KNN graph, with the results shown in Fig.~\ref{ablation5}. We can observe that the use of relatively larger $K$ led to better performance (i.e., comparing $K$=32 and $K$=40 with $K$=16 and $K$=32), as it provides a reasonably large local space that presents more detailed geometric information for the centers. However, too large $K$ (e.g., $K$ = 40 compared with $K$=32) tends to degrade the performance. The reason is that the Euclidean distance fails to approximate geodesic distance when $K$ is too large, thereby destroying the geometry of each patch~\cite{spatial1}. Besides, a very large $K$ means that the central node is more likely to aggregate geometric feature from neighbors that belong to different segmentation class. This is harmful for the network to learn discriminative information for each central node. 
Therefore, we chose $K$ = 32 in the implementation of our TSGCN.

\subsection{Limitations}
Although our TSGCN achieves the leading performance in the task of 3D dental segmentation, it still has certain limitations in handling exceptional cases with 12 teeth. Specifically, for a 12-teeth intra-oral scanner image, our TSGCN may generate false prediction on T6. This can be interpreted by the fact that the outermost tooth of the 12-teeth intra-oral scanner images is annotated as T6, which is usually annotated as T7 in the normal intra-oral scanner image. To address this issue, including more 12-teeth cases as training samples should be considered in our future work.

\section{Conclusion}
\label{conclusion}
A two-stream network, called TSGCN, has been proposed in this paper to segment individual teeth from the intra-oral scanner images acquired by intra-oral scanners. Our TSGCN first applies two input-aware graph learning streams to extract high-level single-view geometric features from coordinates and normal vectors, respectively. Then, it further adopts a self-attention-based feature-fusion branch to combine the complementary information from the two heterogeneous views, by which discriminative multi-view feature representations can be learned for precise cell-wise segmentation. An extensive comparison has been performed for our TSGCN and other five state-of-the art methods on a real-patient dataset. The results demonstrate the superiority of our proposed method, especially for the practically challenging cases.

\textbf{Acknowledgment} This work is supported by the National Natural Science Foundation of China (No. 62176035, 61906025, 82101058, 11690011, U1811461), Chongqing Research Program of Basic Research and Frontier Technology (No. cstc2020jcyj-msxmX0835, cstc2021jcyj-bsh0155, cstc2020jcyj-msxmX0525), the Science and Technology Research Program of Chongqing Municipal Education Commission under Grant (No. KJZD-K202100606, KJQN201900607, KJQN202000647, KJQN202100646). Chongqing Yuzhong District Basic Research and Frontier Exploration Project (No. 20200117). Key Project of Smart Medicine of Chongqing Medical University (No. ZHYX202101)

{\small
\bibliographystyle{ieee.bst}
\bibliography{main}
}

\end{document}